\documentclass[aps,prl,twocolumn,showpacs]{revtex4}
\usepackage[dvips]{graphicx}
\usepackage{amssymb}

\begin{document}

\title{Monte Carlo simulated dynamical magnetization of single-chain magnets}

\author{Jun Li}
\author{Bang-Gui Liu}
\email[Corresponding author:~]{bgliu@iphy.ac.cn}
\affiliation{Beijing National Laboratory for Condensed Matter
Physics, Institute of Physics, Chinese Academy of Sciences,
Beijing 100190, China}

\date{\today }

\begin{abstract}
Here, a dynamical Monte-Carlo (DMC) method is used to study
temperature-dependent dynamical magnetization of famous Mn$_2$Ni
system as typical example of single-chain magnets with strong
magnetic anisotropy. Simulated magnetization curves are in good
agreement with experimental results under typical temperatures and
sweeping rates, and simulated coercive fields as functions of
temperature are also consistent with experimental curves. Further
analysis indicates that the magnetization reversal is determined by
both thermal-activated effects and quantum spin tunnelings. These
can help explore basic properties and applications of such important
magnetic systems.
\end{abstract}

\maketitle

Various nanoscale spin chains have been attracting great attention
because of their important properties and potential applications in
information science and
technology\cite{loth,hein,scm1,scm2,scmrev1}. The single-chain
magnet (SCM) is a new member of such nanoscale spin chains, and its
basic spin unit come from some transition-metal or rare-earth ions
combined with appropriate organic
molecules\cite{hein,scmrev1,scm1,scm2,scm2a,scm3,scm4,scm5,scm6,scm7,scm8,scm9,scm10}.
A famous SCM is the [Mn$_{2}$Ni]
system\cite{scm1,scm2,scm2a,scm3,scmrev1}, with
C$_{62}$H$_{64}$N$_{10}$O$_{14}$Cl$_2$Mn$_2$Ni and
C$_{60}$H$_{66}$N$_{12}$O$_{14}$Cl$_2$Mn$_2$Ni as two typical
formula units with spin $S=3$. A well-known Arrhenius law has been
observed for their spin relaxation at high enough temperature
\cite{arr1,arr2}. On the other hand, at low enough temperature,
quantum Landau-Zener (LZ) spin tunneling should play important roles
in their spin reversal\cite{Landau,Zener}. Such phenomena acn be
investigated by using some methods for single-molecule
magnets\cite{smmbook1,smmbook2,lz1,lz2,lz3,lz4,lz5,numeric1,numeric2,lgb1}.
As for SCM systems, inter-spin exchange interactions play important
roles and thermal effects can cause Glauber spin
dynamics\cite{scmrev1,scm2a}, which was originally proposed for
one-dimensional Ising spin model\cite{glauber,ad1,ad2}. Furthermore,
a systematical experimental study shows that quantum nucleation can
become important to reverse single spins, create domains of reversed
spins, and reverse the whole SCM\cite{scm3}. Therefore, it is useful
to elucidate what roles these play in determining dynamical
magnetization of SCM systems.

Here, we use the hybrid DMC method and thereby investigate the
[Mn$_{2}$Ni] SCM system as a typical example of SCMs, taking both
classical and quantum effects into account. Our results for typical
temperatures and sweeping rates are consistent with corresponding
experimental curves. It is very interesting that we can
satisfactorily fit the simulated and experimental $B_c$-$T$ curves
by one simple function. These means that the DMC method and
simulated results are both reasonable and reliable for such SCM
systems. Furthermore, we explain magnetization reversal modes for
different temperatures on the basis of our simulated results and
analyses. More detailed results will be presented in the following.

The single-chain magnet can be considered a one-dimensional
composite spin lattice whose spins can be constructed by repeating a
basic unit of [Mn$_{2}$Ni]: Mn-Ni-Mn (or
Mn$^{3+}$-Ni$^{2+}$-Mn$^{3+}$). The antiferromagentic Ni-Mn
interaction is much stronger than the ferromagnetic Mn-Mn one so
that the low-temperatures physics of this spin chain can be modelled
by an effective ferromagentic chain of the units of [Mn$_{2}$Ni]
($S=3$) with spin interaction only between the nearest
units\cite{scm1,scm2,scm2a,scm3,scm5,scm6,scm7,scm8,scm9}.

The ferromagnetic spin Hamiltonian can be expressed
as\cite{scm3,scm4,scm6}
\begin{equation}\label{h}
\hat{H} =  \hat{H}_0  - \sum_{i=1}^{N-1} J\hat{\vec{S}}_{i}\cdot
\hat{\vec{S}}_{i+1} - \sum_{i=1}^{N}g\mu _{B}B_{z}\hat{S}_{i}^{z}
\label{e:1}
\end{equation}%
where $g$ is the Lande $g$ factor ($g=2$ is used), $\mu _{B}$ the
Bohr magneton, $J$ ($>0$) the ferromagnetic exchange constant.
$\hat{\vec{S}}_{i}=\{\hat{S}_{i}^{x},\hat{S}_{i}^{y},\hat{S}_{i}^{z}\}$
is the spin vector operator for the $i$-th Mn$_{2}$Ni unit, and
\begin{equation}\label{h0}
\hat{H}_0=\sum_{i=1}^{N} \{-D(\hat{S}_{i}^{z})^{2}-E[(\hat{S}_{i}^{x})^{2}-(\hat{S}_{i}^{y})^{2}] \}%
\label{e:2}
\end{equation}
is the Hamiltonian for the isolated ferromagnetic spin.

$D$ and $E$ are the anisotropic parameters. As for the parameters of
the spin interaction and on-site anisotropy, we use $J/k_{B}$=1.56K
and $D/k_{B}$=2.5K from thermodynamical
measurements\cite{scm3,scm4,scm6}. The transverse anisotropic
parameter $E$ is much smaller, but necessary to realize the
Landau-Zener spin tunnelling. We take $E/k_{B}$=0.1K by comparing
our simulated results with experimental ones.

We use a dynamical Monte Carlo method to simulate the spin dynamics
of the interacting spin system under sweeping magnetic
field\cite{ly1,lbg,lgb1}. At the beginning, we set all of the spins
at the state $S^z=-3$.

We divide the time $t$ into small time steps with a step length
$\Delta t$ and describe the Monte Carlo time points with $t(n)$,
where $n$ takes 0, 1, 2, 3,.... The magnetic field starts from
$-B_{0}$ and increases by an increment of $\Delta t\cdot \nu$ until
$B_{0}$. The spin can be reversed within a Monte Carlo step (MCS)
through the two reversal mechanisms.

\begin{figure}[!tbp]
\includegraphics[width=7.2cm]{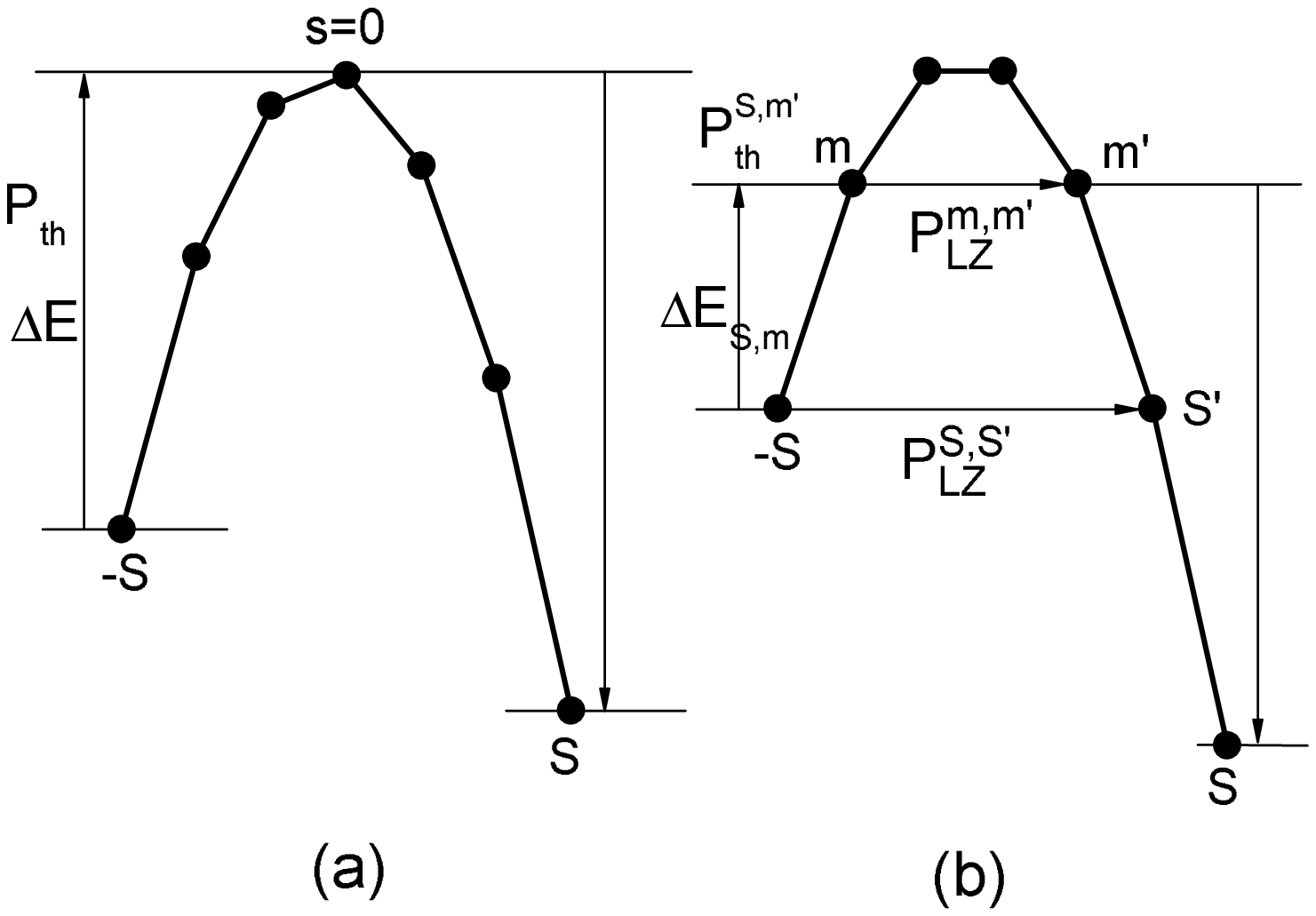}
\caption{ A schematic of the three spin reversal mechanisms:
Thermal-activated barrier hurdling (a), direct and thermal-assisted
LZ tunnelings (b). The horizontal solid line with arrow means that
the two energy levels satisfy the resonance conditions. The
probabilities, energy levels, barrier, and other symbols are defined
in the text.} \label{fig:1}
\end{figure}

For the classical thermal activation, we can obtain the following
probability $P_{th}$ that within the time decrement $\Delta
t$\cite{arr1,arr2}.
\begin{equation}
P_{th}=1-\exp (-R_i\Delta t) \label{e:3}
\end{equation}
Where $R_i=R_{0}\exp (\frac{-\Delta E_i}{k_{B}T})$ is the transition
rate, $k_B$ is the Boltzmann constant, $T$ temperature, and $R_{0}$
the characteristic frequency for the spin system (3$\times
$10$^{8}$s$^{-1}$). $\Delta E_i$ is the potential barrier of the
$i$-th spin between $S_i^z=-3$ and $S_i^z=3$, as shown in Fig.
\ref{fig:1}.

There is a necessary condition for a LZ tunnelling of a spin to
occur: one of the spin energy levels on the side must be equivalent
to another, for example, $E_{m}(t)=E_{m^\prime}(t)$, as shown in
Fig. \ref{fig:1}. With the neighboring spins taken into account,
such conditions are satisfied at the given magnetic
fields\cite{lgb1}. The corresponding LZ transition probability is
given by
\begin{equation}
P_{LZ}^{m,m^\prime}=1-\exp \left[ -\frac{\pi (\Delta _{m,m^{\prime
}}^{{}})^{2}}{2\hbar g\mu _{B}|m-m^{\prime }|\nu }\right]
\label{e:4}
\end{equation}%

where the tunnelling splitting $\Delta _{mm^\prime}$ is the energy
gap at the avoided crossing of states $m$ and $m^\prime$, and $\nu $
denotes the sweeping rate of the magnetic field.

When $m$ equals to $S=-3$ and $m^\prime$ is $S^\prime$, we obtain a
direct LZ tunnelling with the probability
$P_{LZ}^{d}=P_{LZ}^{S,S^\prime}$. For other possible LZ tunnelling
to happen, the spin at first must be excited from $S^z=-3$ to the
$m$ values through some thermal activations, as shown in Fig. 1.
Considering the thermal probability $P_{th}^{S,m}$ which can be
obtained by using the expressions (\ref{e:3}), the probability of
spin reversal in this channel, $P_{LZ}^{m}$, is given by $P_{LZ}^{m}
=P_{th}^{S,m}\cdot P_{LZ}^{m,m^\prime}$. All the three spin-reversal
channels are combined to give the total probability for a spin
reversal\cite{lgb1}:
\begin{equation}
P^{tot}=1-\left( 1-P_{th}\right) \cdot \left( 1-P_{LZ}^{d}\right)
\cdot \prod_m \left( 1-P_{LZ}^{m}\right) \label{e:5}
\end{equation}

In our simulations, we take $\Delta t=0.1$ms and use 100 units of
Mn$_{2}$Ni with free boundary condition. The magnetization is
calculated by averaging $S^z_i$ over the 100 spin sites. Each data
point is calculated by averaging 10000 independent runs to reduce
possible errors. The value of $B_{0}$ is made large enough to obtain
complete hysteresis loops with the help of a symmetrization
treatment.

\begin{figure}[!tbp]
\includegraphics[width=7.5cm]{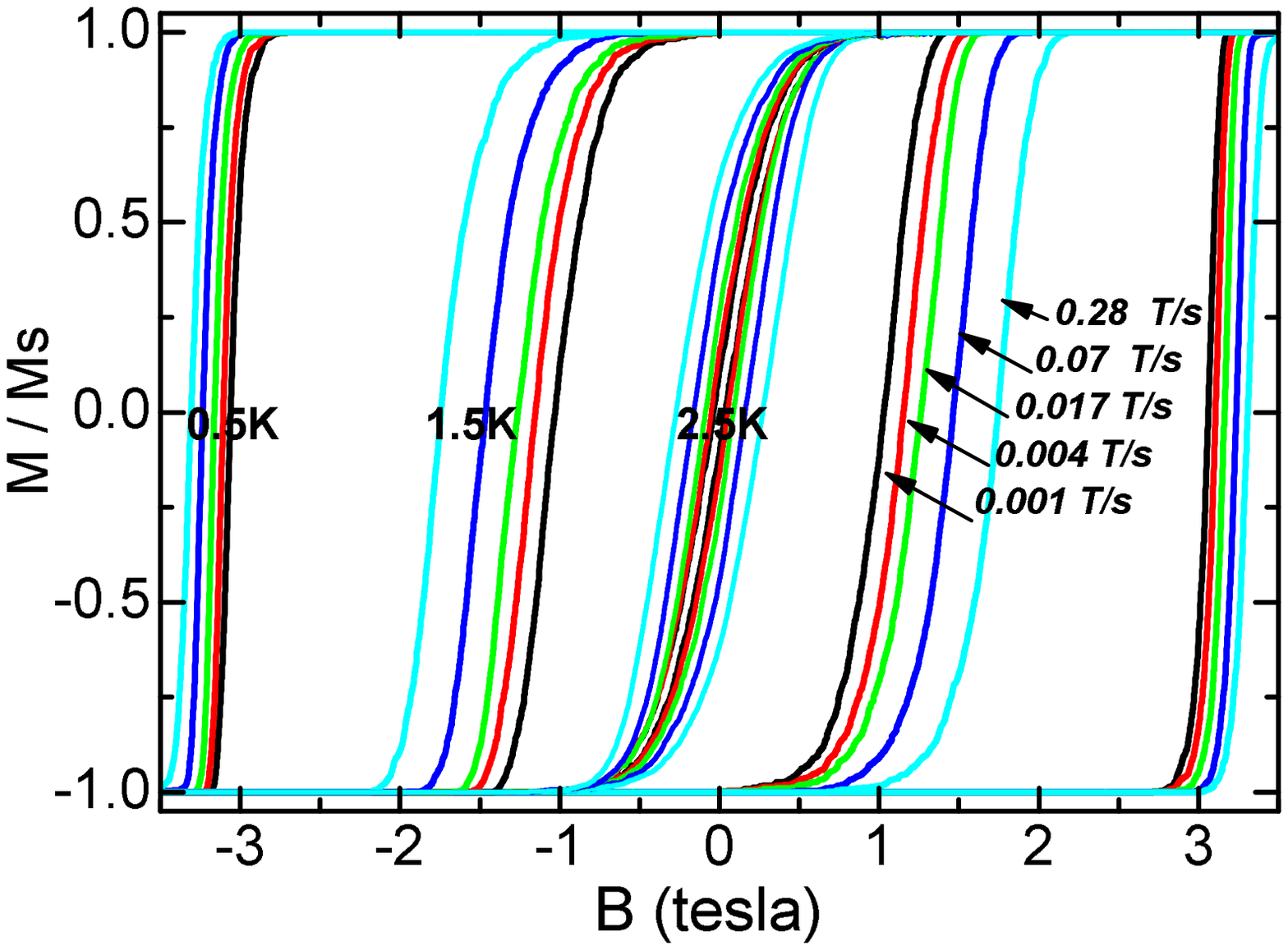}
\caption{Hysteresis loops (normalized magnetization ($M/M_s$) curves
against the sweeping field $B$) for three temperatures: 0.5, 1.5,
and 2.5 K. For every temperature, five magnetization curves are
plotted with five field sweeping rates: 0.001, 0.004, 0.017, 0.07,
and 0.28 tesla/s (from the innermost loop to the outermost for each
temperature).} \label{fig:2}
\end{figure}

Presented in Fig. \ref{fig:2} are our typical simulated
magnetization curves for five different field sweeping rates $\nu$
(0.001, 0.004, 0.017, 0.07, and 0.28 tesla/s) at three different
temperatures $T$: 2.5, 1.5, and 0.5 K. The simulated results show
that the hysteresis loops are strongly dependent on both temperature
$T$ and field sweeping rate $\nu$. Our simulation shows that there
is no hysteresis loop for all the field sweeping rates when
temperature reaches 3 K, and at 2.5 K, the thermal effects are
dominant and spins can be easily reversed, which results in very
small hysteresis loops. Our data analysis indicates that when the
temperature further decreases, the thermal-activated spin reversal
becomes less important and the thermal-assisted LZ spin tunnelling
already takes place frequently. At 1.5 K, another typical
temperature, these two channels are available for the spin being
reversed, but the total reversal probability is less than that of
2.5 K, and hence the coercive fields is substantially larger than
that of 2.5 K. When the temperature becomes very low, for example
down to 0.5 K, our probability analysis reveals that the thermal
activation is almost frozen and the spin reversal can be realized
only through the direct LZ spin tunnelling, and as a result, the
coercive fields are large because the transverse parameter $E$ is
very small. Even at this low temperature, there is no clear step
structure in the magnetization curves, which should be attributed to
the strong spin exchange interaction in the $J$-term. This is in
contrast to those in the cases of Mn$_{12}$ and Fe$_{8}$
systems\cite{smmbook1,smmbook2,lgb1}. Our simulated magnetization
curves also show that the larger the field sweeping rate, the larger
the hysteresis loop. This trend can be explained by considering that
larger sweeping rate means shorter time for spins to try towards
reversal, as shown in classical nanoscale spin
systems\cite{ly1,lbg}.

\begin{figure}[!tbp]
\includegraphics[width=7.2cm]{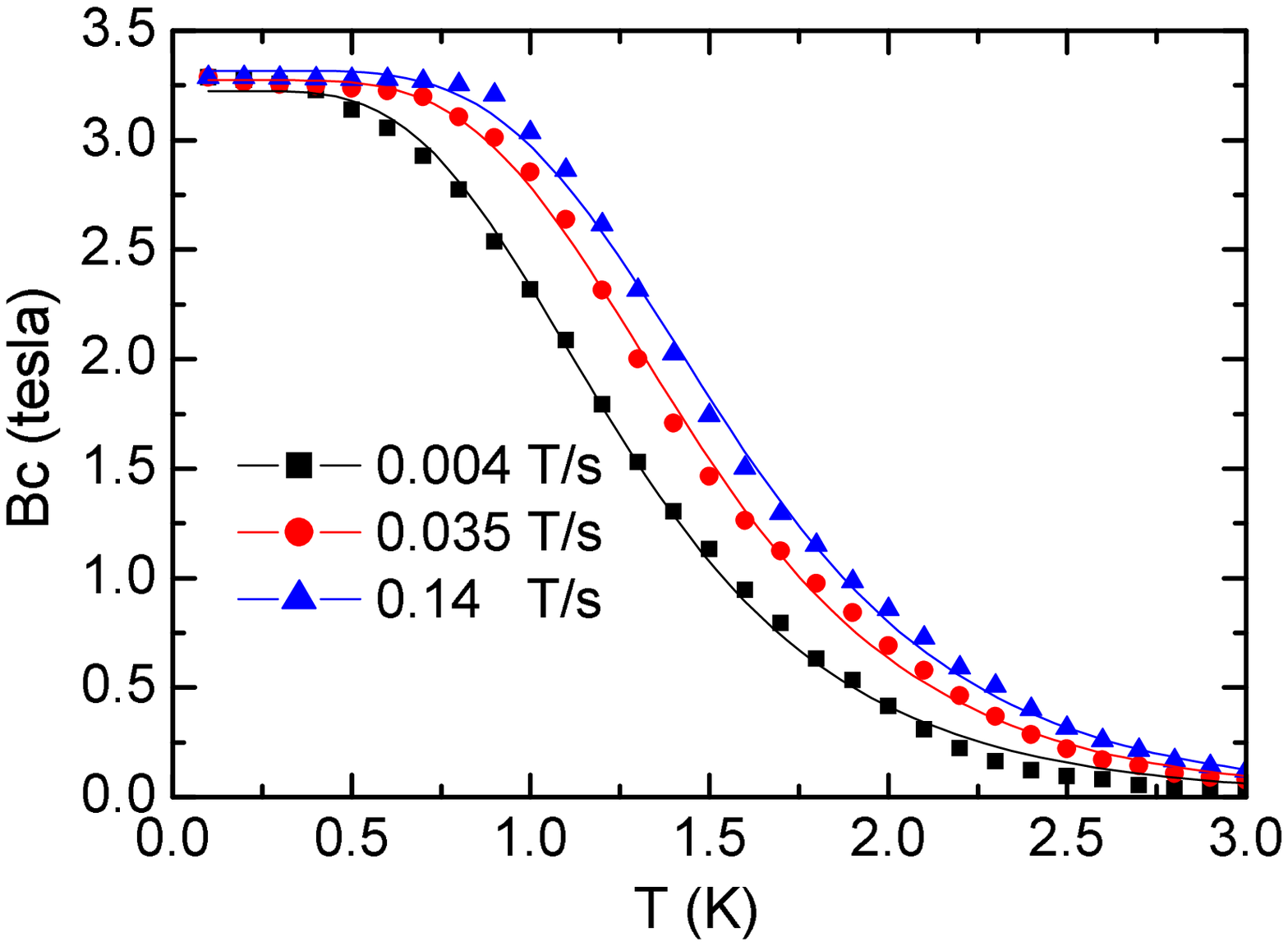}
\caption{Temperature ($T$) dependence of the coercive fields ($B_c$)
for the three field sweeping rates: 0.004, 0.035, and 0.14 tesla/s.
The curves are well fitted with Eq. (\ref{e:6}) using the set of
parameters ($A$,$T_1$,$T_2$) presented in Table  \ref{tab:1}.}
\label{fig:3}
\end{figure}

Furthermore, we have done more simulations with more field sweeping
rates and more temperatures. In Fig. \ref{fig:3} we present our
systematical results on the coercive fields $B_c$ as functions of
temperature $T$ for three sweeping rates $\nu$: 0.004, 0.035, and
0.14 tesla/s. For all the three field sweeping rates, it is clear
that the coercive fields decrease with temperature increasing. It is
very interesting that these $B_c$-$T$ curves  can be well fitted by
the following simple function.
\begin{equation}
B_{c}=\frac{A}{\displaystyle 1+\exp ( \frac{T}{T_1}-\frac{T_2}{T})}
\label{e:6}
\end{equation}
For the three $B_c$-$T$ curves in Fig. \ref{fig:3}, the fitting
parameters ($A$, $T_{1}$, $T_2$) are summarized in Table I. When the
temperature is high, the $B_c$-$T$ curves are dominated by the $T_1$
term in the exponential, $B_{c}\sim A \exp (-\frac{T}{T_1})$, which
should be naturally attributed to thermal activations. When the
temperature is below 1 K, the coercive fields substantially deviate
from classical behavior. Especially when the temperature decreases
below 0.5 K, the coercive fields tend to saturate, $B_{c}\sim A $.
This means that the low-temperature saturation behavior is
consistent with quantum LZ effect, in contrast with Glauber
dynamics\cite{glauber,ad1,ad2}.

It is very surprising that the simple function (\ref{e:6}) can
satisfactorily describe the experimental $B_c$-$T$ curves for such
sweeping rates, too. Our fitted parameters for the experimental
curves are summarized in Table II.

\begin{table}[!htb]
\caption{Fitting parameters of the three theoretical $B_c$-$T$
curves in Fig. \ref{fig:3} in terms of the function defined in Eq.
(\ref{e:6}).} \label{tab:1}
\begin{ruledtabular}
\begin{tabular}{llll}
 $\nu$ (tesla/s) &   $A$ (tesla) &  $T_1$ (K) & $T_2$ (K)\\ \hline
 0.004 &   3.22  &  0.62 &  2.54 \\
 0.035 &   3.27  &  0.65 &  3.27 \\
 0.14  &   3.31  &  0.67 &  3.63 \\
\end{tabular}
\end{ruledtabular}
\end{table}

In high-temperature region, the magnetization reversal is
characterized by easy classical end-site nucleation and fast
classical wall-moving growth of the reversed spin domain. At
intermediate temperature such as 1.5 K, the magnetization reversal
is realized by many-site quantum nucleations and classical
wall-moving growth of the reversed spin domains. In the
low-temperature region, the magnetization reversal is due to
frequent many-site quantum nucleations of the reversed-spin domains
and these domains are effectively merged by subsequent spin
tunnelings. Importantly, it can leads to crossover between these
three modes to change temperature. Therefore, the three modes have
already been unified into one mechanism in terms of our theory (our
model treatment plus our simulation).

\begin{table}[!htb]
\caption{Fitting parameters of the three
experimental\cite{scm3,scm4,scm6} $B_c$-$T$ curves in terms of the
function defined in Eq. (\ref{e:6}).} \label{tab:3}
\begin{ruledtabular}
\begin{tabular}{llll}
 $\nu$ (tesla/s) &   $A$ (tesla) &  $T_1$ (K) & $T_2$ (K)\\ \hline
 0.004 &   2.63  &  0.63 &  2.85 \\
 0.035 &   2.82  &  0.70 &  2.99 \\
 0.14  &   2.97  &  0.75 &  3.08 \\
\end{tabular}
\end{ruledtabular}
\end{table}

As are clearly shown in Table \ref{tab:3}, our theoretical
parameters are in good agreement with those from experimental
data\cite{scm3,scm4,scm6}. In addition, we consider a
three-dimensional spin system by introducing a very weak inter-chain
spin exchange coupling. Our Monte Carlo simulation indicates that
its sublattice magnetization as a function of temperature is
consistent with experimental results concerned\cite{scm8}. Our
simulated magnetization curves shown Fig. \ref{fig:2} and $B_c$-$T$
curves in Fig. \ref{fig:3} are both in good agreement with
experimental curves\cite{scm3}. These show that our model treatment
and simulation methods are reliable and our simulated results, with
parameters from experiment, are reasonable.

As for spin dynamics in SCM systems, Glauber behavior, usually with
some modifications due to finite size effects, is frequently
observed, and on the other hand, there are convincing evidences that
quantum nucleation plays some important roles in the magnetization
dynamics. Our simulated results show that both the classical thermal
activation and quantum spin tunneling play important roles in
determining the spin dynamics. For high temperatures, the classical
thermal activation is dominating, but at very low temperatures the
classical effect becomes less important, even is frozen, so that the
spin dynamics is determined mainly by the quantum spin tunneling
effect.

In summary, we have made the hybrid DMC method suitable to studying
the spin dynamics of SCMs with strong magnetic anisotropy, and used
it to investigate temperature-dependent dynamical magnetization
behaviors of the famous [Mn$_{2}$Ni] SCM system. Our DMC simulated
magnetization curves are in good agreement with experimental results
under typical temperatures and sweeping rates. We have also
calculated the coercive fields as functions of temperature and
plotted $B_c$-$T$ curves for typical sweeping rates. It is
interesting and surprising that our simulated $B_c$-$T$ curves are
well consistent with experimental ones, and both of the simulated
and experimental curves can be satisfactorily fitted with the simple
function in Equ. (6). These means that our theory and simulated
results are reasonable and reliable to other SCM systems and those
made from adatoms on surfaces\cite{loth,hein,new}.

\begin{acknowledgments}
This work is supported by Nature Science Foun- dation of China
(Grant No. 11174359), by Chinese Department of Science and
Technology (Grant No. 2012CB932302), and by the Strategic Priority
Research Program of the Chinese Academy of Sciences (Grant No.
XDB07000000).
\end{acknowledgments}

\end{document}